\documentstyle[12pt,aasms4]{article}

\begin{document}

\title{Abundance Analyses of Field RV Tauri Stars, V: 
 DS Aquarii, UY Arae, TW Camelopardis, BT Librae, U Monocerotis, 
 TT Ophiuchi, R Scuti, and RV Tauri }

\author{
Sunetra Giridhar}
\affil{Indian Institute of Astrophysics;
Bangalore,  560034 India\\
giridhar@iiap.ernet.in}

\author{David L.\ Lambert }
\affil{Department of Astronomy; University of
Texas; Austin, TX 78712-1083
\\ dll@astro.as.utexas.edu}

\author{Guillermo Gonzalez\footnote{Visiting Observer, Cerro Tololo
Inter-American Observatory, which is operated by the Association of
Universities for Research in Astronomy, Inc. under contract
with the US National Science foundation.}}
\affil{Department of Astronomy;
University of Washington;
Seattle, WA 98195-1580
\\gonzalez@astro.washington.edu}

\begin{abstract}

Abundance analyses are presented and discussed for eight RV Tauri variables.
The RV\,B star UY\,Ara  shows the abundance anomalies seen in other RV\,B
stars, namely elements that condense into grains at high temperature are
underabundant but elements of low condensation temperature are much
less underabundant. This pattern is ascribed to a separation of dust from gas
with accretion of gas but not dust by the atmosphere. Abundances for two
RV\,C stars with  earlier results for other RV\,C stars show
that these intrinsically
 metal-poor stars do not show effects of dust-gas separation.
Analyses of five RV\,A stars show that these cooler stars are very
largely unaffected by dust-gas separation. It is proposed that the
deeper convective envelope of cooler stars dilutes anomalies resulting
from dust-gas separation. Possible sites for dust formation and
dust-gas separation - the dusty wind off the RV\,Tauri variable or
a dusty circumbinary disk - are reviewed and observational tests suggested. 

{\it Subject headings: stars:abundances -- stars:AGB and post-AGB --
stars: variables:other (RV\,Tauri)}

\end{abstract}

\section{Introduction}

In this series of papers, we have been exploring the chemical compositions of
RV Tauri variables in the galactic field. This exploration continues
here with abundance analyses reported for seven additional variables
and an improved analysis provided for an eighth star.   Beginning
with the first paper in the series, an analysis of the southern
variable IW\,Car (Giridhar, Rao, \& Lambert 1994), we have shown that
the atmospheric composition of a  RV\,Tauri star may be  abnormal  in the
sense that those elements that condense at ``high'' temperatures ($\sim
1500$ K) into dust grains are underabundant (Gonzalez, Lambert \& Giridhar 1997a,
Paper II; Gonzalez, Lambert \& Giridhar 1997b, Paper III; Giridhar, Lambert
\& Gonzalez 1998, Paper IV). A likely interpretation
is that the atmosphere accreted gas but not dust  from a site of dust
formation. 
%Presence of strong infra-red excesses for some but not
%all RV Tauri variables shows that dust is present near the stars.
%Radiation pressure from starlight may drive dust grains through the
%gas which, if then accreted by the star, will alter the atmosphere's
%composition.

The severity of the atmospheric abundance anomalies differs from
one RV Tauri to another. By enlarging the sample, we hoped to
ascertain the principal factors influencing the effects of a
dust-gas separation on the atmospheric composition. 
Several clues have been provided:
\begin{itemize}
\item
The severity of the abundance abnormalities
is not correlated with the magnitude of the infra-red excess (Paper IV).
\item
Variables with an intrinsic metallicity [Fe/H] $\lesssim -1$, as
assessed from their S and Zn abundances, are not subject to effects
of a dust-gas separation. This conclusion is confirmed by
abundance analyses of similar variables in globular clusters (Gonzalez
 \& Lambert 1997; Russell 1997, 1998; Carney, Fry \& Gonzalez 1998).
\item
Post-AGB stars exhibiting extreme effects of dust-gas separation
are spectroscopic binaries in which the separation may occur in
the circumbinary disk (Waters et al. 1992; Van Winckel, Waelkens \&
Waters 1995). 
Van Winckel et al. (1998) have shown that 
AC\,Her, a well studied RV Tauri variable  exhibiting effects of 
dust-gas separation (see also Giridhar, Lambert, \& Gonzalez 1998; 
Klochkova \& Panchuk 1998),
is a spectroscopic
binary of long period. Van Winckel et al. (1999) argue that direct and
indirect evidence supports 
a claim that ``binarity is a widespread phenomenon
in the RV Tauri class of objects'', and that the separation occurs in the
circumbinary disk and not the stellar wind off the RV\,Tauri star.  
\end{itemize}

 These  clues to the intriguing
physics behind dust-gas separation affecting some but not all
RV Tauri variables are  considered afresh and extended
 in this paper in which we present
abundances for the eight variables 
listed in Table 1 with pertinent data including 
the spectroscopic class (RV A, B, or C) as introduced by Preston et al. (1963)
 and the photometric class (RV a or b).
The RV\,C stars have weak metal lines and a high radial velocity (Joy 1952),
and, hence, may be presumed to be intrinsically metal-poor. The RV\,A and RV\,B
stars belong to the disk population. The RV\,B stars are generally of earlier
spectral type than the RV\,A. Relative to MKK standards, the
metal lines appear weak in the RV\,B but not in the RV\,A stars.   
Six of the present sample are RV\,A stars and triple the number of RV\,A
stars in our survey.
 RV\,b and RV\,a variables differ  in that the
characteristic light curve of alternating deep and shallow
minima with a period of 50 - 150 days  is modulated on a long timescale
($\sim$ 600-2600 days) in the case of the RV\,b stars.

\section{Observations and Abundance Analyses}

Our abundance analysis uses  the observations summarized
in Table 2. A majority of the stars were observed with
 the McDonald Observatory's 2.7m Harlan J. Smith reflector with the
CCD-equipped `2dcoud\'{e}' spectrograph (Tull et al. 1995). A spectral
resolving power of 60000 was used and a broad spectral range was covered
in a single exposure.
The star UY Ara, as was BT\,Lib,  was observed with the CTIO 4m telescope
and a Cassegrain echelle spectrograph at a resolving power of 20,000.

Several stars were observed on  more than one occasion. 
Spectra were rejected
if they showed line doubling, markedly asymmetric lines, or
strong emission at H$\beta$. (Emission was almost always present
at H$\alpha$.) It is presumed that the spectra not
showing these characteristics represent the atmosphere at its
stablest time when standard theoretical models may be applicable.
This presumption should be tested by analysis of a series of spectra taken
over the pulsational cycle. This remains to be done but in
previous papers we have analysed several stars using spectra taken at
different phases and obtained consistent results.
A striking example was given in Paper III where three observations
of SS\,Gem gave widely different effective temperatures (4750, 5500, and 6500 K)
but similar results for the composition.
The abundance analysis was performed exactly as described in
earlier papers of this series.

\section{The Chemical Compositions}

Three signatures may be looked for in the chemical compositions: (i) the 
initial composition, (ii) the effects of deep mixing during stellar evolution
on the initial composition, and (iii) the effects of the dust-gas separation.

The  initial
 composition may be
anticipated from published studies on less evolved stars that show
a generally smooth run of [X/Fe] with [Fe/H], at least over the metallicity
range appropriate for field RV Tauri variables - see 
Lambert (1989), Wheeler, Sneden \& Truran (1989), Edvardsson et al. (1993),
and McWilliam (1997).\footnote{Usual spectroscopic notation is adopted: [X/Y]
= log$_{10}$(X/Y)$_{star}$ - log$_{10}$(X/Y)$_{\odot}$.}

Deep mixing affects the light elements (primarily C, N, and O with, perhaps, a
modest effect on Na) and the heavy or $s$-process elements. Although 
the evolutionary origins of RV Tauri variables are unknown in detail,
it may be assumed that they experienced the first dredge-up that brings
CN-cycled material into the atmosphere. This reduces the C abundance
and increases the N abundance. If RV Tauri variables have  evolved from
the AGB on which they may have experienced thermal pulses and the
third dredge-up, they are expected to be enriched in C  and the $s$-process
elements.

Dust-gas separation more severely depletes those elements
of high condensation temperature  (e.g., Ca, Ti, and Fe)
than those of low condensation temperature
 (e.g., S and Zn, as well as C, N, and O).  A correlation between
the abundance [X/H]  and the calculated condensation temperature
T$_c$ is taken as evidence for dust-gas separation. Here, we adopt
values of T$_c$ given by Lodders \& Fegley (1998) for cooling gas of solar
composition  in  equilibrium at a pressure of 10$^{-4}$ bar. These
estimates are in general in good agreement with the values computed by Wasson
(1985) used in our previous papers. (The largest relevant difference
is for Si for which T$_c^{LF}$ = 1529 K but T$_c^W$ = 1311 K.)  
Even were dust-gas separation the paramount effect in determining the
atmospheric composition, we would not expect a tight correlation
between [X/H] and T$_c$ for the obvious reasons that dust formation
near a star
is unlikely to occur through equilibrium condensation, and the composition
of the gas will, in general, not be solar. 

In the following sections, we present and discuss the compositions
of our eight variables and an additional three stars with published
abundance analyses. 
Relations between
abundance anomalies and infra-red excesses and other observables
are given later when a review is made of the full sample of
21 RV Tauri variables.

\subsection{The RV C Star DS Aquarii}

DS Aquarii was featured in
Paper IV. The present analysis is based on a new spectrum that
provides broader spectral coverage and, hence,  more absorption
lines.  
Metal lines are few and weak in the spectrum, as  expected of a
RV C star. The Fe\,{\sc i} and Fe\,{\sc ii} lines provide atmospheric
parameters (Table 2) quite similar to those derived earlier;
the iron abundance of [Fe/H] = -1.1 is within 0.1 dex of the
previous value. 

DS\,Aquarii's composition (Table 3) is  that
of a normal metal-poor star.
 With the exception of K and Ba,
our new analysis confirms previous results with differences between the analyses
falling within the 
range of +0.3 dex to -0.2 dex. Abundances are provided for the first
time for C, Mg, Cr, Zn, and Eu. One characteristic of a normal
[Fe/H] = -1 star is an overabundance of the $\alpha$-elements. DS\,Aqr
has this overabundance: [$\alpha$/Fe]  = 0.4 (Mg), 0.4 (Si), 0.3 (S),
and 0.2 (Ca) for an unweighted mean of 0.3, which is typical of
normal stars.
 The relative overabundance of Eu ([Eu/Fe] = 0.5) from a single
Eu\,{\sc ii} line is also typical (Woolf, Tomkin, \& Lambert 1996).
 Potassium,
 previously reported overabundant, is here shown to have a normal
abundance ([K/Fe] = 0.3 from the 7699\AA\ resonance line) according
to sparse data on K abundances in metal-poor stars (Gratton \&
Sneden 1987); Ba, also previously reported overabundant,
has an  abundance ([Ba/Fe] = 0.2) that indicates
little to no enhancement. 

The O abundance derived  from the [O\,{\sc i}] 6363\AA\ line
is in good agreement with our earlier result based on the
O\,{\sc i} 7772-5\AA\ triplet: the result [O/Fe] = 0.8 is consistent
with recent measurements from the O\,{\sc i}7772-5 \AA\ and OH
lines in subdwarfs (Israelian et al. 1988; Boesgaard et al. 1999).
The carbon
abundance  given by the C\,{\sc i} lines at 9078, 9088, and 9111\AA\
is [C/Fe] = 0.15. This is comparable to the carbon abundance
of subdwarfs as measured by Gustafsson et al. (1999) from the
[C\,{\sc i}] 8727\AA\ line. The apparently slightly lower C
abundance of DS\,Aqr is roughly consistent with expectations for the
first dredge-up that the star must have experienced. 
The atmosphere is C-poor in the sense that C/O $\simeq$ 0.05.

Signatures of a dust-gas separation are absent with  S, Sc, and Zn in
particular, having normal abundances:
 [S/Fe] = 0.3 is typical for this $\alpha$-element in
a [Fe/H] = -1 star, and [Sc/Fe] = 0.0 is also a normal
result.  Zinc, not previously measured, is normal for this metallicity
(Sneden \& Crocker 1988).

\subsection{The RV\,B  Star UY Arae}

Lloyd Evans (1985) assigned this star to the spectroscopic class
RV\,B. It is not known to which photometric class (RVa or RVb) UY\,Ara
belongs.
Our abundance analysis from a 
CTIO Cassegrain spectrum was hampered by the lack of unblended
lines. Derived atmospheric parameters are given in Table 2 and
abundances in Table 3.

Inspection  shows that the dust-gas separation
has affected the atmosphere; a clear indicator is the
high relative abundances of S and Zn ([S/Fe] = 1.0, and [Zn/Fe] =
0.7) and the low relative abundance of Sc ([Sc/Fe] = -0.7). 
Sulphur and Zn, if completely unaffected by the dust-gas separation,
imply an intrinsic metallicity of [Fe/H] $\simeq$ -0.3. There is a clear
correlation (Figure 1) between the abundances and the condensation
temperature (or interstellar depletion). 
(Although the oxygen abundance was not measured, the atmosphere is
probably O-rich: C/O $< 1$.)

\subsection{The RV A Star TW Camelopardis}

Preston  et al. (1963) assigned the spectroscopic classification
RV\,A to this  RVa  star whose light curve more closely resembles
that of a  Cepheid than a textbook RV\,Tauri variable.
Our abundance analysis (Table 3) is from an excellent {\it 2dcoud\'{e}}
spectrum which provides the atmospheric parameters summarized in
Table\,2. 

The iron abundance is [Fe/H] = -0.5. Relative to this
value, most elements have the  abundance expected of such a
mildly metal-poor star, i.e, dust-gas separation has not
imprinted a clear signature on the composition. In particular,
S and Z
have roughly normal abundances ([S/Fe] = 0.5 and [Zn/Fe] = 0.1)
as do Ca, Sc, Ti, and Al with higher condensation temperatures
than Fe ([Ca/Fe] = -0.1, 
[Sc/Fe] = 0.0, [Ti/Fe] = -0.1, and [Al/Fe] = 0.1). The
S abundance is marginally higher than normal.
Abundances of the several heavy elements
from Y to Eu  are  normal. A mild overabundance is indicated
for $\alpha$-elements Si, Mg, and S but not for Ca and Ti, which may
reflect the fact that the spectrum is rich in lines.

Carbon provides the only abundance anomaly: [C/Fe] = 0.7 from a 
set of 6 C\,{\sc i} lines. The O abundance from the [O\,{\sc i}]
lines
is as expected for a mildly metal-poor star. This analysis
shows the star to be C-rich with C/O $\simeq$ 2.

\subsection{The RV C Star BT Librae}

Remarkably few observations of any kind
 have been reported for this high-velocity
star. The period of 75.3 days is traceable to Ashbrook (1942).
Our abundance analysis is based on two spectra: a CTIO Cassegrain echelle
spectrum of 5300-8400\AA\ and a McDonald  2dcoud\'{e} spectrum.
The derived atmospheric parameters are identical to within the errors
of measurement. 
The heliocentric radial velocity was +140 $\pm 2$ km s$^{-1}$ on
1998 June 6 and +136 $\pm$ 2 km s$^{-1}$ on 1998 July 31. Although
the classification RV\,C has not been previously given, we
feel justified in assigning it based on the radial velocity and
the paucity of lines in the spectra.

The composition (Table 3) is in all respects that of a normal  star
of [Fe/H] = -1.2. The $\alpha$-elements show the expected overabundance:
[$\alpha$/Fe] = 0.3 (Mg), 0.5 (Si), 0.3 (S), 0.3 (Ca), and 0.3 (Ti)
for an unweighted mean of 0.3 dex.
Dust-gas separation has not affected this star; the abundances of
S, Ca, Sc, and Zn  relative to Fe are all normal.

Oxygen based on the [O\,{\sc i}] 6300\AA\ and 6363\AA\ lines is
overabundant, [O/Fe] = 0.9, by about the degree found for DS\,Aqr,
and consistent with recent measurements for subdwarfs (Israelian et al.
1988; Boesgaard et al. 1999).\footnote{These recent measurements are
based on the O\,{\sc i} 7772-5\AA\ lines and the OH ultraviolet
lines. Measurements of the O abundance in subdwarfs to low
metallicities from the [O\,{\sc i}] lines are difficult to impossible
for subdwarfs because the line is very weak. Our measurements
using the forbidden lines that are probably formed under LTE
 in these supergiants give support
to these controversial recent measurements from the O\,{\sc i} and OH
lines.} 
Carbon (C\,{\sc i}) lines were not detected or [C/Fe] $\leq$ -0.1
and 
C/O $\leq$ 0.05.

\subsection{The RV A Star U Monocerotis}

This star is a fine example of a RV b photometric variable. 
 Percy et al.
(1997) give the principal period as 92.23 days and the long period as
2475 days. Pollard \& Cottrell (1995) from a radial velocity study
find the star to be a spectroscopic binary with an orbital period
consistent with that of the long-term photometric variability. 

Our abundance analysis is summarised in Table 4 where published
analyses by Luck \& Bond (1989) and Klochkova \& Panchuk (1998)
are also tabulated.  There is agreement about the metallicity with
the three analyses providing [Fe/H] = -0.7 to -0.8.
The abundance differences between our analysis and Luck \& Bond's,
 $\delta$(X) =
 [X/H]$_{us}$ - [X/H]$_{LB}$,  are small:
 $\delta$(X) is
 in the range $\pm$ 0.2 for a
majority of the elements in common and the largest difference is
-0.5 for Al. The agreement with Klochkova \& Panchuk is
also satisfactory;
the corresponding $\delta$(X)  is less than $\pm$ 0.3
 with the exception of 5 elements (S, Mn, V, La, and Sm)
In view of this pleasing consistency, we average the abundances
for the purposes of discussion; all major conclusions  are valid
for our analysis. 

To within $\pm$ 0.2 or 0.3 dex, the relative abundances [X/Fe] for all
but two elements are
those expected for a normal star with [Fe/H] $\simeq$ -0.6. Sodium is
possibly overabundant, [Na/Fe] = 0.7. Carbon at [C/Fe] = 0.7 is
overabundant but oxygen with [O/Fe] = 0.5 is approximately normal for
a metal-poor star; the
atmosphere appears to be O-rich with  C/O $\simeq$ 0.8.

In several key respects, U\,Mon resembles AC\,Her: both are RVb
photometric variables,  long-period spectroscopic binaries, and 
stars with a pronounced infra-red excess. They differ in that AC\,Her
has a strong signature of dust-gas separation (Paper IV; Van Winckel et al.
1998)
 but U\,Mon does not.
They differ too in effective temperature, dereddened B - V color, 
and spectroscopic class (RV A for U\,Mon but RV\,B for AC\,Her). 

 To contrast the lack of a dust-gas
signature in U\,Mon  and the strong signature in AC\,Her (Paper IV),
 we compare elemental abundances. Note that their intrinsic
metallicities are equal to within the errors of measurement:
[Fe/H] = -0.7 for U\,Mon versus the inferred initial [Fe/H] = -0.8
for AC\,Her.
 The abundance differences 
between U\,Mon and AC\,Her are essentially independent of condensation
temperature  for T$_c \leq 1600$ K but increase
sharply for elements of higher T$_c$. If $\Delta$[X/H] =
[X/H]$_U$ - [X/H]$_{AC}$ where [X/H]$_U$ is the mean abundance for
U\,Mon and [X/H]$_{AC}$ is the abundance for AC\,Her reported in
Paper IV, we find $\Delta$[X/H] = 0.41 $\pm 0.15$ from 10 elements
from Mg to Nd with T$_c$ of 1300 K to 1600K.
 Five elements with T$_c \geq$ 1600 K give
$\Delta$[X/H] from 0.7 for Sc to 1.9 for Al. At the other extreme,
S and Zn with T$_c \simeq 680$ K give $\Delta$[X/H] of 0.3
and 0.2 dex respectively.

\subsection{The RV A Star TT Ophiuchi}

 Our spectrum of TT\,Oph
indicates that this is one of the cooler variables in the
program. 
This RV\,Tauri variable has no detectable infra-red
excess. 
 Derived atmospheric parameters are listed in Table 2,
and abundances in Table 3.

TT\,Oph's
atmosphere  most probably has  the   composition expected  for a
 star
of the derived metallicity [Fe/H] = - 0.8. Zinc with [Zn/Fe] = 0.0
and potassium with [K/Fe] = 0.1
suggest that, if dust-gas separation has been active, it has not
affected elements with condensation temperatures cooler than
that of iron. 
Elements with the highest condensation temperatures of the sample
may indicate effects of dust-gas separation: Ca, Sc, and Ti
are slightly underabundant relative to normal expectation. 
The observed  values are [Ca/Fe] = -0.3, [Sc/Fe] = -0.3, [Ti/Fe] = 0.0, 
but expected values are approximately 0.2, 0.0, and 0.2,
respectively where differences between observation and
expectation are
about the error of measurement.

  One might contrast TT\,Oph
and DS\,Aqr (and BT\,Lib). In the latter cases, the expected signature
of a positive value of [$\alpha$/Fe]  was clear. It is not as uniformly
seen in TT\,Oph where [$\alpha$/Fe] =  0.4 (Si), 0.6 (S), -0.3 (Ca), and
0.0 (Ti). We suspect that this difference is largely a reflection of
the contrasting spectra. DS\,Aqr and BT\,Lib, which are  warmer
and  more metal-poor than TT\,Oph, have cleaner spectra
that allow for more certain line identifications and more accurate
measurements of equivalent widths than  in the case of TT\,Oph.

One abundance anomaly is present, however, for TT\,Oph. Elements
heavier than zinc are consistently 
underabundant: [X/Fe] = -0.3 (Y), -0.7 (Ce),
-0.5 (Pr), -0.5 (Nd), -0.3 (Sm) for elements with 3 or more lines. 
These elements have condensation temperatures higher than that of
iron so that the underabundance might reflect a dust-gas separation, as
was hinted at from the Al, Ca, Sc, and Ti.

\subsection{The RV A Star R Scuti}

This RV\,Tauri variable has been
extensively observed. An abundance analysis of a spectrum obtained
near the secondary light maximum was described by Luck (1981). 
The 
reported metallicity [Fe/H] = -0.9 was  not exceptional but the
low abundance of $s$-process elements deservedly drew Luck's
attention: [$s$/Fe] $\simeq -1$ is not expected for a mildly metal-poor
star.  Our analysis is based on a  spectrum at a phase
similar to Luck's observation.
Derived atmospheric parameters are listed in Table 2 and abundances
in Table 5.

The metallicity  [Fe/H]  = -0.4 is higher than the -0.9 found
by Luck but equal to Preston's (1962) value from a curve of
growth analysis. Elements from C to Zn have a normal abundance,
[X/Fe] $\simeq 0.0$, with one exception: scandium with [Sc/Fe] = -1.1
is greatly underabundant. Although the Sc abundance is based on
just 3 Sc\,{\sc ii} lines, it is clear that the abundance [Sc/Fe] = 0
is excluded by our analysis. Then, either Sc is really underabundant
or the Sc\,{\sc ii} lines are weakened by  anomalous excitation. 
Scandium is, of course, the element in our sample with the next to
highest condensation temperature (T$_c$ = 1652 K). Aluminum with
T$_c$ = 1670 K is only slightly underabundant with [Al/Fe] = -0.4.
 Titanium with T$_c$ = 1600 K
has a normal abundance ([Ti/Fe] = -0.1). Calcium with T$_c$ = 1634 K
is slightly underabundant: [Ca/Fe] = -0.2 according to Luck (1981) who
did not measure the Al or Sc abundances but also found [Ti/Fe] = -0.1.
There is no strict requirement or confident expectation that anomalies should
correlate extremely well with T$_c$. Nonetheless, where the dust-gas
separation is severe, as in AD\,Aql and AC\,Her (Paper IV), the
anomalies  correlate quite well with T$_c$. If the Sc underabundance is
due to dust-gas separation, the  effects seem restricted to elements
of the very highest condensation temperatures with a dramatic effect
only for Sc.\footnote{According to  Lodders \&
Fegley (1998)  only Al by a margin of 22 K, Hf
 by 38 K, and W by 142 K have higher T$_c$ than Sc.} 

Our analysis confirms Luck's discovery of an  underabundance
of elements heavier than zinc. Figure 2 shows [X/Fe] versus
X's  condensation temperature. As noted above, abundances are
roughly consistent with [X/Fe] = 0 and independent of T$_c$ for
elements from C to Zn with the exception of Sc. 
 Lower abundances and greater element-to-element scatter are
seen for the heavy elements. The scatter may well be due to the
paucity of lines; almost all determinations are based on 1 or 2 lines.
If mean abundances across the sample of measured elements
 are considered, our mean is
higher than Luck's: [X/Fe] = -0.8 from 5 elements versus -1.3 but when
results for the 3 elements in common are averaged the difference  (-0.1 dex)
is
not significant.  There is a hint that the underabundances 
are correlated with T$_c$: Figure 3  in separate panels gives the
abundances from the two analyses with the selection limited to
the heavy elements plus Sc or Ca.
If the Ca abundance is overlooked, Luck's data from
Y to Nd suggests that [X/H] declines with increasing T$_c$.
Perhaps, our 
data also hint at a decline.

 Luck \& Bond (1989) provided a different
explanation for the underabundance of the heavy elements in luminous
stars such as R\,Sct. They suggested that
an intense flux of Lyman photons from shocks results in an
over-ionization of atoms and ions with ionization potentials
of 13.6 eV or less. These heavy elements  exist primarily
as the ions X$^+$ under LTE conditions. Then, the relevant
ionization potential is that of the singly-charged ion.
If a relationship exists between our [X/Fe] and the ionisation
potential of the ion X$^+$, it is less obvious than possible correlations
in Figure 3. In particular, the five heavy elements in our
analysis span a range of 1 dex in [X/Fe], but the ionisation potentials
differ by only 0.5 eV about
a mean of 11.0 eV. One might note that Sc with
the Sc$^+$ ionisation potential at 12.8 eV is as underabundant as
the heavy elements, Eu excepted. Titanium with Ti$^+$'s ionisation
potential of 13.58 eV right at the Lyman edge is, however, not
underabundant.

 Noting that a ratio 
[$s$/Fe] $\simeq -1$ is normal
for extremely metal-poor stars,
 Luck (1981) proposed that R\,Sct was initially an  
extremely metal-poor star which during evolution has become greatly
enriched in helium.
When analysed on the assumption of a normal He abundance,
the star is assigned a much higher abundance [Fe/H]. One expects abundance
ratios such as [$s$/Fe] to be estimated without serious error leading to
the apparent anomaly of a [Fe/H] $\simeq -1$ star with the [$s$/Fe] ratio
of a much more metal-poor star. Since [Sc/Fe]  $\simeq 0$ for normal
stars with [$s$/Fe] $\simeq -1$, our measurement of a drastic Sc underabundance
is in conflict with Luck's ingenious scheme but qualitatively consistent
with the idea of dust-gas separation.

It is known that the atmosphere of R\,Sct (and  RV\,Tauri
variables in general) at certain phases is not well approximated
by a canonical theoretical atmosphere. Preston's (1962) 
study  of spectra obtained over a pulsation cycle showed a line
doubling at some phases. These and other effects have
been interpreted by Gillet et al. (1989) as due to the
passage of shocks through the atmosphere. Our analysis, as was Luck's,
is based on a spectrum in which lines are single and almost
symmetric. 
We largely rely on weak lines that
are probably  formed in the deeper layers and are largely unaffected
by a hot layer above the photosphere. Nonetheless, in light
of the remarkably specific abundance anomalies, it would be
of interest to obtain and analyse spectra across the pulsation
cycle in order to disentangle true abundance anomalies from
atmospheric effects.

\subsection{The RV A Star RV Tauri}

The eponym is 
 one
of the coolest stars which we have analysed.  Atmospheric
parameters derived from our spectrum are listed in Table 2,
and abundances are given in Table 6.

An abundance analysis of RV\,Tauri was reported recently by
Klochkova \& Panchuk (1998). Their results from spectra
obtained at two different phases are also listed in Table 6.
Klochkova
\& Panchuk find an appreciably higher metallicity: [Fe/H] = 0.05
versus our -0.43. 
When [X/Fe] is
considered, the two analyses agree to within $\pm 0.3$ dex for all
elements in common except C (1.0), Sc (0.5), Cr (0.4), and Mn (-0.5)
where the difference in [X/Fe] in the sense of `us - them' is given
in parentheses.  The difference in the carbon abundances is striking.
The oxygen abundances are in good agreement.  Our results imply
RV\,Tau is C-rich (C/O $\simeq 2$) but Klochkova \& Panchuk find it to
be O-rich (C/O $\simeq$ 0.2). 
Of the elements considered by Klochkova \& Panchuk but not by us, one
has an unusual abundance: [S/Fe] = 0.8 where [S/Fe] = 0 is anticipated
for a star with [Fe/H] = 0.1. A high S/Fe ratio is a signature of
dust-gas separation but this is an unlikely explanation in this
case because [Zn/Fe] = 0.1 and, if the S is undepleted, an extraordinary
initial abundance [Fe/H] $\simeq$ 0.8 is implied. Klochkova \& Panchuk
acknowledge the sulphur problem and suppose it reflects non-LTE
effects. A search of our spectrum for S\,{\sc i} lines was
unsuccessful; upper limits to the equivalent widths of the leading
lines give [S/Fe] $\leq $ 1.1. Confirmation of the sulphur abundance
should be sought.

The abundances suggest that RV\,Tau has not been
subject to a dust-gas separation. With few exceptions, elements
 up to and including the
iron peak have normal abundances for [Fe/H]  = -0.4. Notably,
the high-T$_c$ elements  Al, Ca, Sc, Ti  show normal (solar)
ratios relative to Fe to within 0.2 dex. Sodium ([Na/Fe] = 0.6)
appears overabundant, a not unusual result for supergiants.
Zinc with [Zn/Fe] = 0.4 is overabundant  and might indicate
that all elements with T$_c > $ 1200 K are depleted slightly.
However, Klochkova \& Panchuk find [Zn/Fe] = 0.1.
Elements heavier than zinc appear slightly underabundant with a mean
value [X/Fe] = -0.3. Klochkova \& Panchuk's abundances give a similar
underabundance.

\subsection{Other Stars}

Abundances analyses for additional stars have been
reported: AI\,CMi (Luck \& Bond 1989; Klochkova \& Panchuk 1998)
and RU\,Cen (Luck \& Bond 1989). In addition, two recent analyses
of AC\,Her have been published (Klochkova \& Panchuk 1998;
Van Winckel et al. 1998). Lloyd Evans (1999) has proposed that LR\,Sco be
included among RV\,Tauri variables, a star previously analysed by
Giridhar, Rao \& Lambert (1992).
 These data are reviewed in the light of the
dust-gas scenario.

\subsubsection {AI Canis Minoris}

The iron abundance estimates 
agree to 0.1 dex: [Fe/H] = -1.14 (Klochkova \& Panchuk)
and -1.02 (Luck \& Bond). Abundance ratios [X/Fe] agree to within
0.1 dex in the mean with a scatter of $\pm$ 0.3 about this
value. 
The $\alpha$-elements are slightly overabundant. Judged by
the Al, Ca, Sc, and Ti abundances, the atmosphere is unaffected by
dust-gas separation. The zinc abundance determined solely by
Klochkova \& Panchuk is overabundant with [Zn/Fe] = 0.7 but
it is based on a single line. Klochkova \& Panchuk report
a normal abundance for $s$-process elements: [$s$/Fe] =
0.0 (Y), -0.2 (Zr), 0.0 (Ba), and -0.1 (La) for elements represented
by 2 or more lines, with $r$-process Eu at [Eu/Fe] = 0.3 showing the expected 
slight enrichment (Woolf et al. 1996).  If the same criterion (2 or more
lines) is applied to Luck \& Bond's analysis, the $s$-process
elements appear slightly underabundant: [$s$/Fe] = -0.1 (Zr), -0.4 (La),
and -0.5 (Nd) with Eu at [Eu/Fe] = 0.0.

These analyses are consistent in showing that AI\,CMi has
an intrinsic [Fe/H] = -1.0 and is probably unaffected by dust-gas separation.
We identify it as a RV\,C star. A radial velocity has not been
published. It should be noted that it appears to be substantially
cooler than the RV\,C stars analysed by us; Klochkova \& Panchuk
estimate T$_{\rm eff}$ of 4100 K to 4500 K from three spectra but our RV\,C
stars have T$_{\rm eff}$ in the interval 5300 K to 6500 K.

\subsubsection{RU\,Centauri}

Luck \& Bond (1989) found
[Fe/H] = -1.4. 
The Ca, Sc, and Ti abundances (relative to Fe) appear
normal to within the errors of measurement, as do the two $s$-process
elements Ba and Nd. The star which was reported to have a high O
abundance ([O/Fe] = 1.4) is known for its strong CH  band
(Lloyd Evans 1974). It would be of interest to
determine the C/O ratio.
In light of the fact that several dust-gas affected stars 
show no additional depletion of Ca, Sc, and Ti (relative to Fe)
but markedly anomalous S/Fe and Zn/Fe ratios, we defer a decision
on whether RU\,Cen is  affected until S and Zn abundances are
provided. It is a RV\,B variable and, therefore, almost
certainly  affected - see below.

\subsubsection{AC\,Herculis}

Our analysis of AC\,Her from spectra at two different phases was
given in Paper IV. Two recent papers (Klochkova \& Panchuk 1998;
Van Winckel et al. 1998) provide independent analyses.
Results of all three analyses are summarized in Table 7.
A surprising result is the lack of consistency in the absolute
abundances, for example, [Fe/H] = -1.4 (Paper IV), -0.8 (KP),
and -1.7(VW). It would be of interest to know how much of this
spread is due to different analytical tools and how much to
a failure of the common approach (theoretical static model atmospheres,
assumption of LTE etc.)
when applied to a pulsating star. 
Notwithstanding the differences in absolute abundances, all
three analyses agree that the dust-gas separation is strikingly
evident for AC\,Her. Relative abundances [X/Fe] are in generally
satisfactory good agreement: the mean difference $\delta$(X) between
Paper IV and Van Winckel et al. is a pleasing -0.05 $\pm 0.14$,
and slightly larger with a bigger spread
 between Paper IV and Klochkova \& Panchuk
with the mean  $\delta$(X)  = 0.2 $\pm 0.4$. There is a difference
in the C/O ratio: we and Van Winckel et al.  find the atmosphere
to be O-rich (C/O = 0.4) but Klochkova \& Panchuk find it to be
C-rich (C/O = 1.5).

\subsubsection{LR Scorpii}

An identification of LR\,Sco as a post-AGB star was mooted by Giridhar
et al. (1992) on the basis of its infrared excess and common abundance
anomalies with 89\,Her. In general, LR\,Sco's anomalies are those
we now associate with dust-gas separation:  low Ca and Sc abundances
were noted with [Ca/Fe] = -0.3 and [Sc/Fe] = -0.6 but [S/Fe] = 0.5.
One oddity is the published Zn abundance corresponding to [Zn/Fe] = -0.3
where a value closer to +0.3 would be expected from dust-gas
separation. Possibly, the tabulated Zn abundance is a typographical error.
If the Zn abundance is set aside, LR\,Sco appears to be mildly affected
by dust-gas separation.

\section{Dust-gas Separation - the Requirements}

Atmospheres underabundant in those elements that condense out
of cooling gas into dust grains have now been discovered for
several types of stars: $\lambda$\,Bo\"{o}tis (main sequence)
stars (Venn \& Lambert 1990),
 the post-AGB stars (Venn \& Lambert 1990; Bond 1991; Van Winckel,
Mathis \& Waelkens 1992), and, most
recently, ST\,Pup, a W\,Vir or Type II Cepheid variable,
 (Gonzalez \& Wallerstein 1996)\footnote{Lloyd Evans (1983) in a report on
Type\,II Cepheids noted the similarity of ST\,Pup's spectrum to that of the
RV\,B variables. One presumes that the other Type\,II Cepheids are not
affected by dust-gas separation.}, and
several RV\,Tauri variables. It is likely that the
RV\,Tauri variables evolve into post-AGB stars but unlikely that
all post-AGB stars have evolved from RV\,Tauri variables (Jura 1986).
The $\lambda$ Boo stars are a quite unrelated class of stars. 

The basic idea that is referred to as a dust-gas separation is
readily described as a four-step process: dust grains
condense out of cooling gas at
a site exterior to the stellar atmosphere, a `phase separation' of
dust from gas occurs at the site, gas from the site
  is accreted by the stellar atmosphere, and
the accreted gas remains in the atmosphere largely unmixed with
the envelope below the atmosphere.
One may identify some
general requirements to produce an observable
abundance anomaly and broad conditions under which the requirements
may be satisfied:

\begin{itemize}
\item Requirement $A$.
Accretion by a stellar atmosphere of gas of anomalous composition  will
result in
observable abundance anomalies  provided that the accreted gas
is  not diluted through convection
or other processes with  a much larger sub-atmospheric
reservoir of normal composition. 

Qualitatively, this condition is met by the
$\lambda$ Boo and post-AGB stars.
 The RV Tauri variables span a range in effective temperature.
The hottest examples, F-type supergiants,
 overlap the cooler post-AGB stars. The coolest
examples are late K to early M supergiants with presumably  substantial
outer convective envelopes.

\item Requirement $B$.
After dust grains have formed and been separated at least partially
from the  gas, 
the star must accrete gas. The efficiency of this accretion process
is likely to depend on the relationship between the affected star
and the source of the cooling gas in which the dust-gas separation
occurs.

Two broad possibilities may be noted:
\begin{itemize}
\item
The star affected by dust-gas separation may be  both the
source of the cooling gas and the receiver of the cleansed gas. 
A pertinent example is a  star with 
dust-gas separation occurring in its stellar wind. 
In order to achieve the required return of
gas (not dust) to the atmosphere, either the outflow must be  intermittent
at least near the base of the wind or 
asymmetric with outflow over parts of the surface with inflow of gas occurring
elsewhere from layers affected by
dust-gas separation.

\item
The cooling gas is exterior to the affected star which accretes
gas.
 An example might
be a binary system in which gas is accreted from a circumbinary disk
containing material  ejected  from one or other of the
stars in the binary. If the ejection occurred long previously,
 accretion
of this gas suffices to create abundance anomalies on the post-AGB star
at least up to the time that the reservoir is exhausted.
\end{itemize}

\item Requirement $C$.
A mixture of dust and gas near a star is subject to radiation
pressure with the dust grains experiencing by far the greater
force. Thus, the separation of dust from gas is encouraged by 
the radiation pressure  on the grains and discouraged by
the drag of the gas on the grains.  It is reasonable to assume that
the star with the anomalous abundances is the source of the
radiation pressure. The luminosity of this star is a factor
promoting  dust-gas separation.
Since hydrogen and helium atoms outnumber all other species
by a large margin, the  drag on a grain is independent of the
metallicity of the cooling gas, i.e., the metallicity of the
star.\footnote{Earlier, we implied that
the drag on a grain in  low metallicity gas was higher than in
high metallicity gas because of a less favorable dust to gas ratio
 (Gonzalez \& Lambert 1997). We thank Ruth 
Knill-Ngani for pointing out our error.}

\item Requirement $D$.
A requirement that dust form in a region from which gas may
be subsequently accreted by the star is 
transparently obvious. Its implications, however,
may be fundamental to understanding the key observables on
which the observed abundance anomalies may depend. In its
crudest form, the requirement means that the timescale for
dust formation must be shorter than the time for which the gas
is cold. Interpretation of this condition depends on the physical
conditions experienced by the gas. For example, consider two
identical parcels of gas with one maintained at constant temperature
and density, and the other subjected to periodic shocks propagating 
out from a pulsating star's atmosphere. It is clear that requirement
$C$ will lead to different constraints on physical conditions and gas
composition, and, in the second example, on pulsation properties.  
Metallicity will
influence the time required to assemble a grain; this time at
a fixed density and temperature will scale approximately as
M$^{-2}$ where M is the abundance of a grain constituent. In
locations where physical conditions are changing, inadequate
grain formation may occur in low metallicity gas to sustain
a dust-gas separation. Under constant physical conditions, as, perhaps,
 in a circumbinary disk, the limiting
metallicity may be lower.

\end{itemize}

\section{Dust-Gas Separation - Observations and Correlations}

Our previous papers in this series suggested that the abundance
anomalies arising from dust-gas separation were common but not
universal among RV\,Tauri variables. This paper adds a new
dimension to the circumstances under which anomalies are
absent. In this section, we present the observational clues
to the basics of dust-gas separation. Pertinent data on 21
stars given in Table 8 include the spectroscopic and photometric
type, the pulsation period,
 the unreddened mean B-V color, selected abundances as well
as the inferred initial iron abundance [Fe/H]$_0$ (see Paper IV),
the C/O ratio.

It must be recognized that not only the
scale of the anomalies differs from one star to the next but
the pattern of the anomaly may differ. 
We introduce an index DG (dust-gas) to denote five gradations of the
abundance anomalies:
\begin{itemize}
\item {\bf DG0}: The abundance anomalies are absent: DG0 stars
include the RV\,C stars (e.g., DS\,Aqr) and a majority of the
observed RV\,A stars (e.g., TW\,Cam).
\item {\bf DG1}: Abundance anomalies are confined to  the
elements with the highest predicted T$_c$, i.e., Al, Ca, Sc, 
and Ti. Our results suggest that Sc is generally more
depleted than the others and Ca and
Ti may be normal (relative to Fe) when Sc is quite severely
underabundant.  Our prototypical DG1 star is R\,Sct. Possible
members include DY\,Aql and CE\,Vir for which the abundance
data is incomplete at present, particularly lacking  are
the S, Ti, and Zn abundances for DY\,Aql, and the S, and Ca abundances
for CE\,Vir.
\item {\bf DG2}: The abundance anomalies extend to iron and
elements of a similar lower T$_c$ than Sc with
no measureable difference in magnitude, i.e., [Sc/Fe] $\simeq$ 0.
Representatives of this group are EP\,Lyr and AC\,Her.
\item {\bf DG3}: The abundance anomalies decrease with decreasing
T$_c$ , i.e., [S/Fe] $>$ 0 and [Sc/Fe] $<$ 0. Examples of  DG3
stars are SS\,Gem and UY\,Ara.
\item {\bf DG4}: The pattern of abundance anomalies is that of a
DG3 star but more severe, i.e., [S/Sc] $>$ 2 for DG4. Examples of
DG4 stars are IW\,Car and AR\,Pup.
\end{itemize}

The index assigned to individual stars is given in Table 8.

\subsection{Infra-red Excesses}

A simple model of dust-gas separation would predict a correlation
between the abundance anomalies and
 an infra-red excess that measures the circumstellar
dust. 
Near-infrared excesses were measured and discussed by Lloyd
Evans (1985) and Goldsmith et al. (1987). In Figure 4,
we present the 2-color diagram J-K vs K-L based on their
measurements. Stars are identified by name and
distinguished  according to the
presence (filled circle denotes DG1, 2, 3, and 4 stars)
or absence (open circle denotes DG0 stars) of abundance anomalies.
(A finer distinction between  DG1, 2, 3, and 4 is uninformative,
perhaps because the sample sizes are small.)  A few stars are not
plotted for lack of measurements. The presence
or absence of an infra-red excess can be judged from the
J-H vs H-K diagram for three additional stars: AD\,Aql has a very slight
excess, and DS\,Aqr and V453\,Oph have the colors of unreddened
stars.
Three stars - V360\,Cyg, EP\,Lyr and CE\,Vir -
lack published JHKL photometry.
Infra-red excesses detected from JHKL photometry arise from
warm dust, i.e., dust close to the star. For stars (UY\,Ara, RU\,Cen,
AC\,Her, BT\,Lib, and R\,Sge) without a
substantial excess at longer wavelengths, Goldsmith et al. estimated
dust shell radii in the range of 4 to 12 stellar radii and dust
temperatures of 600 to 2000 K.

Spectral energy distributions published by Goldsmith et al. 
show that the principal flux excess of some RV\,Tauri stars
occurs at longer wavelengths indicating the presence of
cooler dust.  Gehrz (1972) and Gehrz \& Ney (1972) reported
11.3 $\mu$m photometry for many of our stars. We put the IRAS 12$\mu$m fluxes
on the Gehrz scale to provide results for additional stars.
Data for 16 of our 21 stars  are shown in the
J-K vs J -[11.3] diagram (Figure 5). 

Inspection of Figures 4 and 5 suggests one firm conclusion: 
an  infra-red excess is not a guarantee that the
star is affected by dust-gas separation.  At J-K $\lesssim
1.5$ and J-[11.3] $\lesssim$ 6.5, unaffected (DG0) and affected
(DG1-4) stars are intermixed:  dusty
RV\,Tau and U\,Mon show   no convincing evidence for dust-gas
separation, and  AD\,Aql and SS\,Gem with
striking abundance anomalies show  little evidence for a dusty
shell. It is true that the reddest trio of stars in both figures
are severely affected by dust-gas separation so perhaps there
is a critical infra-red excess above which dust-gas separation
is necessarily pronounced. 

 Infra-red excess is an incomplete measure of the total
presence of dust; the strength of the excess depends on the 
chosen wavelengths  or, equivalently, on the temperature
distribution of the dust. This is well shown by inspection of the
IRAS colors (Raveendran 1989): ranked by the [12] - [25] index,
the reddest five objects in our sample in order of decreasing
index are DY\,Ori, AC\,Her, IW\,Car,
CT\,Ori, and RV\,Tau but ranked by K-L color the reddest five
are AR\,Pup, IW\,Car, UY\,Ara, CT\,Ori, and RV\,Tau.
 It appears that the presence of an infra-red excess is
not a sufficient requirement for an atmosphere to show effects
of dust-gas separation. 
 The reddest
object at [12] - [25] is the RV\,B star RU\,Cen, a run of the mill object in
J - [11.3] (= 4.9) and K-L (=0.50). Unfortunately, too little 
is known presently about its composition to determine if dust-gas
separation has occurred. It probably has as all RV\,B stars 
in our sample are affected.

\subsection{Effective Temperature}

Inspection of Table 8 shows that {\it all} RV\,B stars are
affected by dust-gas separation but only 2 (R\,Sge and SS\,Gem)  of the RV\,A
stars are markedly affected. This dichotomy  in large measure was 
anticipated by  the spectroscopic definition of the RV\,B stars: blue spectra
show strong bands of CH and CN but  the metal lines  are weak relative to
MKK standards.  In contrast the RV\,A stars show strong metal lines.
A second 
difference may hold a key to understanding  the dust-gas separation: the
RV\,A stars are generally of later spectral type and so cooler than the
RV\,B stars. Lloyd Evans (1974) provides a clear statement of the defining
spectroscopic characteristics of the RV\,A, B, and C stars.

A simple measure of the difference between RV\,A and RV\,B is provided from the 
temperatures derived from our spectra. Recall that
spectra were selected for analysis only if the lines were
single and symmetric, a condition met only over a particular
phase range in the pulsation. The derived temperatures for
the RV\,B stars are clearly warmer than for the RV\,As. 
A temperature T$_{\rm eff} <5000$ K is common for the
RV\,As  but the minimum temperature recorded for a RV\,B is 4750 K
at one phase for SS\,Gem with two other observations providing
5500 K and 6500K. The difference in T$_{\rm eff}$ between the
two classes may be fundamental for controlling the dust-gas
separation (see discussion of requirement $A$ above) but not
entirely responsible for the differences in their spectra to which
the differences in composition arising from dust-gas separation
must be a contributor.

 Temperatures measured for   the field 
RV\,C stars, which show no evidence for dust-gas
separation, are in the same range as the RV\,B stars with the exception of
the cool RV\,C AI\,CMi.  
Of the variables in globular clusters\footnote{Zsoldos (1998) has 
argued that the light curves of the half-dozen stars in globular clusters
classified as RV Tau variables fail the photometric criteria of this 
class (as determined from the field RV Tau's), in particular the lack of
alternating minima. While Zsoldos does make useful points concerning the
real differences between the light curves of of field and globular RV Tau's,
we do not believe the differences are sufficient justification to remove 
them from the RV Tau class. Variations in the minima of globular cluster
RV Tau's are in fact seen in the light curves (e.g., Martin 1938, 1940)
and velocities (Wallerstein 1958). We suggest a new subcategory
specific to globular cluster RV Tau's, RV(g).}, three of the five
analysed by Gonzalez \& Lambert (1997)  and V1 in $\omega$ Cen analysed
by Gonzalez \& Wallerstein (1994) are in the temperature range of the
field RV\,C and RV\,B stars. Two may be cooler. Only V1 shows any signs
of gas-dust separation, with its low [Al/Fe]. The range of [Fe/H] among 
$\omega$ Cen giants is quite large, -2.0 to -0.5 (Norris \& DaCosta 1995
and references therein), so it is possible V1's original metallicity
was much higher than its present [Fe/H] = -1.8. It seems safe to claim
that the absence of dust-gas separation in the  RV\,C
(and some of the cluster)  variables  and in the RV\,A variables
occurs for different reasons.

\subsection{Chemical Composition}

One striking correlation between abundance anomalies and
metallicity  noted earlier is confirmed here by the
analyses of BT\,Lib, and V453\,Oph. RV\,C variables do
not show effects of dust-gas separation. Their composition
is that expected of a star having the derived Fe abundance,
i.e., $\alpha$-elements are slightly enhanced. The fact that
they are high-velocity objects also implies that they are
intrinsically metal-poor.  Similar variables in globular
clusters confirm this result; the variables have
 the composition of red giants of their host
cluster. 
Our small sample of RV\,C stars spans the metallicities
[Fe/H]  = -2.2 to -1.1. The warmer globular cluster stars
are from clusters with [Fe/H] = -1.2 and -1.5.  
Metallicity not effective temperature
appears to be the key parameter separating the immune RV\,C stars
from the affected RV\,B stars, and distinguishing the RV\,C stars
from the similarly immune RV\,A stars.  

In discussing the metallicity of the RV\,B stars, the question
arises as to what is the relevant metallicity: the measured
[Fe/H] (or other abundance such as [Si/H] or [Mg/H]) or 
the [Fe/H] inferred from the (assumed) undepleted [S/H] and [Zn/H], a quantity
designated [Fe/H]$_0$ in Table 8 (see Paper IV)?
The measured [Fe/H] range from -0.9 to -2.3 with [Fe/H]$_0$
in the range 0.2 to -0.7. On noting that the lower limit
for [Fe/H]$_0$ is similar to the upper limit for the RV\,C stars,
we could speculate that [Fe/H]$_0$ $\simeq$ -0.8 is a critical
metallicity: dust-gas separation is effective for stars
with {\it initial} [Fe/H] $>$ -1.0 but ineffective for
more metal-poor stars. 

Two possible  ways to account for the critical [Fe/H] come to mind.
Suppose that the pulsation controlling the stellar wind of the
RV\,Tauri variable is  determined in part by the metallicity of the envelope,
that is likely to be unaffected by dust-gas separation.
If dust-gas separation occurs in the
wind, it is the initial metallicity that 
(indirectly) determines   the separation process.
The critical [Fe/H]  implies then that pulsational properties and
their influence on dust-gas separation in the winds is weak for
an initial [Fe/H] $<$ -1
and strong for [Fe/H] $>$ -1.  A quite different example of control
is possible if the dust-gas separation site is remote from the star, as in
the case of gas ejected earlier by the star and recaptured. In the case of
a binary,  gas having the initial composition may be ejected by 
either star and stored in the
circumbinary disk.  If the metallicity is low, gas may escape the
storage site before dust grains can form and dust-separation can begin.
For neither of these examples is it yet possible to justify that
the critical metallicity is the empirical limit of [Fe/H] $\simeq -1$.

The C/O ratio (Table 8) of the RV\,A and RV\,B stars indicates that the
great majority of the stars are O-rich with a mean C/O of
about 0.4 (the solar ratio) with a suspicion that C/O is higher
for the RV\,A stars and, hence, for the DG0 and DG1 stars. Three
stars are C-rich, the DG0 stars TW\,Cam and R\,Sge, and the DG4 star
IW\,Car. The RV\,C stars are very C-poor with C/O $\sim 0.05$.
To within the measurement errors, the C/O ratios of the O-rich
stars are consistent with the idea that the stars have not
experienced C enrichment through evolution on the AGB or 
by other means; the C/O ratios including those of the RV\,C stars
are roughly consistent with initial abundances and a reduction of
C at the first dredge-up.

Despite the few elements analysed, often one element per star, it
is clear  that the stars are not enriched in the $s$-process
elements. Enrichment is assessed by comparing abundances of $s$-process
elements and abundances of iron-group elements of the same
T$_c$. Table 8 gives the estimated [$s$/X] which in some cases
are revised from those in Paper IV. Not one star is
convincingly enriched in  $s$-process products. Certainly, none
exhibit the enrichment typical ([$s$/X] $\sim 1$) of AGB stars that
have experienced the third dredge-up. Unless there is a subtle
systematic error in the abundance analyses or in the interpretation
in terms of T$_c$, we must conclude that RV\,Tauri variables are
not descended from thermally-pulsing AGB stars.

\subsection{Binarity}

All severely
affected ([Fe/H] $\lesssim -3$) post-AGB stars are in binary
systems (Van Winckel et al. 1995), as are some of the less affected
stars. What are thought to be single post-AGB stars are less
affected by dust-gas separation. In addition, it is argued that
some but not all of the binary post-AGB stars present evidence for a  warm
dusty disk (cf. Van Winckel et al. 1999) that
was suggested by Waters et al. to be the site for the dust-gas
separation.
  (One star - BD+39$^{\circ}$\,4926 - is well known for
its lack of an infra-red excess!)

This suggestion motivated Van Winckel et al. (1999) to search for
a (causal) link between RV\,Tauri variables affected by dust-gas
separation and binarity.
  Spectroscopic binary orbits have
been determined for just three stars: AC\,Her (Van Winckel et al. 1998),
U\,Mon (Pollard \& Cottrell 1995), and EN\,TrA (Van Winckel et al. 1999.)
AC\,Her is seriously affected by dust-gas separation but the RV\,A U\,Mon is 
unaffected. The composition of  EN\,TrA  is unknown.
Other stars certainly show long-term
velocity variations indicative of orbital motion: good examples
are IW\,Car (Pollard et al.
1997), and EP\,Lyr (Paper II), both greatly affected by dust-gas 
separation.

Various proposals have been made identifying the long-term
light variations of the RV\,b variables as a consequence of  orbital motion 
(Percy 1993; Fokin 1994; Pollard et al. 1997). 
Van Winckel et al. (1999) extend an
association of RV\,b and binary stars to include the RV\,a
stars by noting that, if
the dust around the stars is primarily located in the
orbital plane, stars viewed at low inclination to the plane
may show  photometric variations with a period equal to the
orbital period. A similar binary viewed at higher inclination  will
be deemed a RV\,a variable.
Van Winckel et al. further adduce evidence, some previously
forwarded by others, in support of a claim that the
dust and gas around RV\,Tauri variables is trapped and stored
near the star, i.e., in a circumbinary disk, and not in an
outflowing wind fed by the star. 

In Van Winckel et al.'s scenario,  the RV\,Tauri stars affected by dust-gas
separation are long-period binaries. Since their assumption
was that  all RV\,A and RV\,B variables are affected, their
bold claim that ``binarity is a widespread phenomenon in the
RV\,Tauri class of objects'' followed.
Although our
present results show  that the RV\,A's are very largely immune, the
proposed scenario is not directly refuted;
the immunity of the RV\,A's may be due to more severe dilution of  accreted gas
by the  thicker convective envelope of a cooler star. The alternative
explanation that RV\,B's are binaries and RV\,A's are not is less
plausible,  not just because the RV\,A U\,Mon is
a spectroscopic binary, but because it is
difficult to understand why
the small difference in effective temperature between the coolest
RV\,B and the warmest RV\,A star is due to the presence of a
companion to the RV\,B star. 
 Van Winckel et al. appear to have
overlooked the RV\,C stars and their immunity to dust-gas
separation.

An extension of Van Winckel et al.'s proposal that `binarity is a widespread
phenomenon' to `binarity is a universal phenomenon' for RV\,Tauri
variables may be rejected on the grounds that it
implies that the stellar pulsations
are controlled by the presence of a companion.
Peculiar stars owing their origin to membership in a binary
are certainly known: the Barium stars are an outstanding example
in which the `peculiar' composition of the Barium star is due to mass
transferred from the companion that evolved first and reached the
AGB before transferring $s$-process and C enriched material. But it is
difficult to accept that the RV\,Tauri's pulsation is controlled
by a distant companion.
There is no evidence that the composition of
RV\,Tauri stars is dominated by mass-transfer to the extent that
the envelope composition that determines pulsational properties
 are determined by  a companion.
 Just conceivably, tidal distortion of the
the star's envelope influences the onset and form of the pulsation.
That theoretical
 models (Tuchman et al. 1993; Fokin 1994) successfully identify
 the pulsations as driven by the partial ionization of the stellar
envelope is evidence that the RV\,Tauri phenomenon is triggered
inside the star and is unlikely to be controlled by an orbiting
companion. 
Discovery of a well defined single period-luminosity law for
RV\,Tauri (and  W\,Virginis) stars of the Large Magellanic Cloud
 (Alcock et al. 1998) suggests that
the RV\,Tauri pulsations cannot be influenced greatly
by a companion; such an influence would be expected to induce scatter in a 
P-L relation. 
Dust-gas separation effects may well be influenced by
binarity but binarity is unlikely to be
 a universal phenomenon for RV\,Tauri stars.

\subsection{Luminosity, Effective Temperature, and Surface Gravity}

A solid claim may now be made from the luminosities of the RV\,Tauri variables
in globular clusters and especially  of those in the Large Magellanic
Cloud (Alcock et al. 1998) that the RV\,Tauri stars are loss mass
objects evolving from red giants. 
The most luminous variables are most probably post-AGB stars evolving
at constant luminosity (Sch\"{o}nberner 1993). RV\,Tauri
stars of short period, W\,Vir, and BL\,Her variables have luminosities
less than that of the tip of the first red giant branch and, therefore,
could be post-RGB or post-AGB. Theoretical calculations (Gingold 1976)
to support a post-AGB interpretation have been reported.

In the absence of accurate distance estimates\footnote{The {\it Hipparcos}
 catalogue has a parallax entered for 8 of the stars in Table 8 but the
errors are large in all cases. In no case
is the quoted error less than 30\% of the measurement.}   field RV\,Tauri
variables, we compare observations and theory through the
log$g$ versus logT$_{\rm eff}$ `HR-diagram' (Figure 6).
Theoretical isochrones from Bertelli et al. (1994) are shown for
three compositions from solar ($Z$ = 0.02) to metal-poor
$Z$ = 0.0004 ([Fe/H] $\simeq$ -1.6)

The  period-luminosity law for RV\,Tauri (and W\,Vir)
stars in the LMC (Alcock et al. 1998) provides accurate estimates
of the stars' luminosities. If the bolometric correction is ignored,
a 75 day RV\,Tauri has log L/L$_{\odot}$ = 3.7, an estimate
that increases by 0.1 to 0.2 dex depending upon the chosen
bolometric correction. The slight metal deficiency of the
LMC is typical of the field RV\,A  stars. Variables in globular
clusters provide a similar luminosity  for metal-poor stars
like the RV\,C variables (Gonzalez \& Lambert 1997). Since the tip of the 
first giant branch is at logL/L$_{\odot}$ = 3.3, these RV\,Tauri stars
may be identified as post-AGB stars rather
 than post-RGB stars. 
Luminosities of our field RV\,Tauri stars
 may be calculated from the atmospheric parameters
T$_{\rm eff}$, log$g$, and an assumed mass (here, 0.8$M_{\odot}$).
The mean value logL/L$_{\odot}$ = 3.4 $\pm$ 0.2  from
17 stars with periods in the range 60 - 95 days is the luminosity
of the tip of the red giant branch.

RV\,Tauri stars are compared in Figure 6 where log$g$ is plotted
versus logT$_{\rm eff}$. For stars observed more than once,
straight means of log $g$ and T$_{\rm eff}$ are plotted.
 In the event that the real stellar atmosphere
is represented adequately by a standard (static) atmosphere, analysis
of the stellar spectrum should return the correct $g$ and T$_{\rm eff}$
and, hence, the actual $L$ subject to an assumption about the stellar
mass. Under such conditions, analyses of spectra obtained at
different phases should fall on  a line of constant $L$ 
(log$g$ = 4logT$_{\rm eff}$ + log$M/L$). 
Multiple analyses of SS\,Gem and R\,Sge satisfy this
requirement  but those of EP\,Lyr and CE\,Vir may indicate a
steeper slope.  Several interesting correlations are revealed in
Figure 6. The RV\,C and RV\,B stars define a single  well-defined 
relation that
is much steeper than a line of constant $L$.  A great majority
of these stars have periods in a narrow range (60 - 95 days) and,
therefore, the relations seems unlikely to be a reflection of the
P-L law directly. The slope could be accounted for if the iso-period
lines in the HR-diagram have positive slope: more luminous stars
being cooler than less luminous stars of the same period.  
Stars in our sample with the shortest (AR\,Pup) and longest (CT\,Ori)
 period are in the middle of the RV\,B distribution in Figure 6.
Metallicity whether it is the initial or the current atmospheric
value would seem not to affect the relation because RV\,B and RV C
stars greatly differ in their metallicity. The temperature
difference between RV\,A and RV\,B stars is striking in Figure 6
with the boundary at T$_{\rm eff} \simeq 5200$ K.

Post-AGB evolution occurs at approximately constant $L$. This
implies that many RV\,Tauri variables evolve off the early-AGB
at a luminosity slightly in excess of log $L/L_{\odot}$ = 3.3.
They are first RV\,A stars\footnote{A phase as a SRd variable
may occur first (Giridhar, Lambert \& Gonzalez 1999).} and evolve
to RV\,B stars developing the signature of dust-gas separation
as the convective envelope disappears with increasing effective
temperature. RV\,Tauri stars of lower luminosity presumably
evolve even earlier from the AGB (or the RGB).  Sch\"{o}nberner (1993)
provides evolutionary tracks for post-AGB stars which run at
constant luminosity to higher T$_{\rm eff}$ from the cool
end of his calculations (log T$_{\rm eff} \simeq 3.8$). The predicted
luminosity runs from log $L/L_{odot}$ = 3.1 to
to 4.2 for  core masses of 0.546$M_{\odot}$ and 0.836$M_{\odot}$ 
respectively. The luminosity spread of the RV Tauri stars is roughly
consistent with the mean mass of about 0.6$M_{\odot}$ for field white dwarfs.

Our demonstration that the $s$-process elements have a normal
abundance would seem to preclude that the stars' predecessors
were AGB stars that experienced the third dredge-up.
 Alcock et al.
show that a single period-luminosity law fits RV\,Tauri variables
and the shorter period W\,Vir or Type\,II Cepheids. We estimate 
logL/L$_{\odot}$ $\simeq$ 2.5 for a variable with a period of 8 days.
These lower luminosities are consistent with  theoretical
proposals that  stars with a low-mass envelope pulsate like RV\,Tauri
variables (Tuchman et al. 1993), and  evolve into the
instability strip  from the AGB to which they may not return (Gingold 1976),
Intriguingly, a low-mass
envelope is a  prequisite for  dust-gas separation to result in observable
abundance anomalies.

\section{Working Hypotheses and Observational Programs}

General requirements for effective contamination of an atmosphere
with gas from a region that experienced a dust-gas separation were stated
earlier. Translation of such requirements into a theory making
quantitative predictions is difficult. As an interim measure and a guide
to observing programs, we advance a few working hypotheses that
tie together the requirements and the observational clues. We offer
two  hypotheses according to whether the key processes
operate in a single star (S) or a binary star (B).

The S hypothesis is as follows:

\begin{itemize}
\item
 Abundance anomalies  present in RV\,B
 stars result from a process of dust-gas separation
in a star's circumstellar shell fed by a wind driven by the
stellar pulsation. 
\item
 The absence of the anomalies in a majority of the RV\,A 
stars is not necessarily due to the failure of  dust-gas separation but
 to inefficient return of gas to the atmopshere and/or to the  dilution of
this gas  by the deeper
convective envelope of these cooler stars.
\item
 The absence of anomalies in the RV\,C stars  is due to the inability
of the dust-gas separation occurring in the stellar wind to
operate efficiently in stars of low intrinsic metallicity.
\item
 The presence of RV\,B stars with  an atmospheric iron
abundance less than that of  some
RV\,C stars  implies that the dust-gas separation is controlled by the abundance of an incondensible element and/or the controlling composition is that of the
deep envelope that retains the star's initial composition.
\end{itemize}

The B hypothesis is as follows:
\begin{itemize}
\item
 The extreme abundance anomalies seen in  post-AGB stars
that are long-period spectroscopic binaries result from dust-gas
separation in material stored in a circumbinary disk of normal
composition from which
gas is accreted by the post-AGB star.
\item
 Binarity is a widespread phenomenon among RV\,Tauri stars.
\item
 Abundance anomalies  of RV\,Tauri stars in 
long-period binaries result primarily from the process that operates in
the post-AGB spectroscopic binaries.
\item
The absence of anomalies in many RV\,A stars is accounted for as in the
S hypothesis. 
\item
 That the most metal-poor RV\,Tauri atmosphere is more
metal-rich by a factor of about 100 than the most metal-poor
post-AGB star is due to a combination of factors: the wind off the 
RV\,Tauri star reduces the rate of accretion from the circumstellar
disk, atmosphere-envelope mixing is more effective for the 
RV\,Tauri
stars, and the abundance anomalies  of the RV\,Tauri stars have
not yet achieved their maximum values due to lack of time.
\item
 The lack of anomalies in RV\,C stars is due to the low metallicity
of the circumbinary gas  that inhibits grain formation and therefore 
the dust-gas separation; gas accreted by the star is not depleted in
condensible elements.
\end{itemize}

The S and B hypotheses may operate in tandem.
It seems improbable that a RV Tauri star must belong to a binary
in order to pulsate in the characteristic manner of the class.
 A determination of the
viability and relative importance of the hypotheses
 should be possible from continued
observational efforts on two fronts: (i) abundance analyses should be extended
to additional RV\,Tauri stars of all spectroscopic classes, and (ii)
a radial velocity program  is needed to test  the idea that spectroscopic
binaries are unusually common among these variables.  The latter test
requires a long-term program to search for small amplitude orbital
changes in the presence of a pulsational change. 
The need for theoretical work on dust-gas 
separation in the context of  the S and B hypotheses is obvious.

Closer scrutiny of the accuracy of the abundance analyses  is
warranted. Particularly instructive would be a study of representative
RV\,A, B, and C stars over a full pulsational cycle. Confidence in
the present standard analysis will be enhanced if it can be
shown that the derived composition is independent of the phase of
the observation except over limited intervals. These spectroscopic
campaigns may also serve to track gas flows out and back into the top
of the stellar photosphere and, perhaps, so to see the dust-gas
separated material flow into the atmosphere.

We are grateful to Katerina Lodders for sending us a table of condensation
temperatures, and to Jocelyn Tomkin for obtaining a spectrum for us.
This research has been supported in part by the Robert A. Welch Foundation of
Houston, Texas and the National Science Foundation (grant AST-9618414).

\clearpage
\clearpage
\section*{Figure Captions}

\figcaption{Abundance [X/H] versus condensation
temperature for UY\,Ara. Elements are identified by their chemical
symbol.  
}

\figcaption{Abundance [X/Fe] versus condensation temperature for R\,Sct.
Filled circles denote our measurements. Open circles denote the
measurements published by Luck (1981). Measurements for the same
element are connected by a line. Elements are identified by their
chemical symbol. The cross represents the Fe abundance [X/Fe] = 0
by definition.
}

\figcaption{Abundance [X/Fe] versus condensation temperature for
R\,Sct and heavy elements. The top panel shows our results including the
measurement of the Sc abundance. The bottom panel shows Luck's measurements
including the measurement of the Ca abundance. The cross represents the
Fe abundance for which [X/Fe] = 0 by definition.
}

\figcaption{Color-color J - K versus K - L plot. Open circles denote
stars unaffected by dust-gas separation (DG0 stars - see text). 
Filled circles denote stars affected by the dust-gas separation (DG1-4
stars - see text). Stars are identified by their GCVS designation
omitting the name of the constellation except that, to avoid confusion,
  DY\,Ori,
 DY\,Aql,  R\,Sge  and
 R\,Sct are noted in full.}

\figcaption{Color-color J - K versus J -  [11.3] plot. Open circles denote
stars unaffected by dust-gas separation (DG0 stars - see text). 
Filled circles denote stars affected by the dust-gas separation (DG1-4
stars - see text). Stars are identified by their GCVS designation
omitting the name of the constellation except that, to avoid confusion,
DY\,Ori, DY\,Aql, R\,Sge, and R\,Sct are noted in full. Three stars are marked
by arrowheads because an upperlimit not a measurement of their [11.3] is
available.}

\figcaption{The log$g$ vs logT$_{\rm eff}$ diagram for RV Tauri variables.
The dashed lines are tracks for a stellar mass of 0.8$M_{\odot}$ evolving 
at the constant luminosity of log$L/L_{\odot}$ = 3.3 (lower line)
and 4.3 (upper line). The theoretical isochrones for an age of 10$^{10}$ years
and compositions $Z$ = 0.0004, 0.004, and 0.02 (solar) include the RGB and
the AGB that begins from the horizontal branch. The RV Tauri stars
are denoted by their Preston spectroscopic class according to the
legend on the figure.}

\end{document}